\documentclass{osa-article}

\journal{oe}
\usepackage[utf8]{inputenc}
\usepackage{color}
\usepackage{bm}
\usepackage{amsmath}
\usepackage{graphicx}
\usepackage{caption,subcaption}
\usepackage{siunitx}
\usepackage{lineno}

\begin{document}

\title{High Efficiency Topological Pumping with\\Discrete Supersymmetry Transformations}

\author{David Viedma,\authormark{*} Gerard Queralt\'{o}, Jordi Mompart, and Ver\`{o}nica Ahufinger}

\address{Departament de F\'{i}sica, Universitat Aut\`{o}noma de Barcelona, E-08193 Bellaterra, Spain}

\email{\authormark{*}david.viedma@uab.cat} %% email address is required



\begin{abstract}
Making use of the isospectrality of Supersymmetry transformations, we propose a general and high-fidelity method to prepare gapped topological modes in discrete systems from a single-site excitation. The method consists of adiabatically connecting two superpartner structures, deforming the input state into the desired mode. We demonstrate the method by pumping topological states of the Su-Schrieffer-Heeger model in an optical waveguide array, where the adiabatic deformation is performed along the propagation direction. We obtain fidelities above $F=0.99$ for a wide range of coupling strengths when pumping edge and interface states.
\end{abstract}

%%%%%%%%%%%%%%%%%%%%%%%%%%  body  %%%%%%%%%%%%%%%%%%%%%%%%%%
\section{Introduction}

Due to their general robustness against defects and disorder, topological states have gathered widespread attention in several fields \cite{Hasan2010,Lu2014,Yang2015,Huber2016,Sato2017}. For optical devices, harnessing the unique properties of these states could lead to a variety of applications, such as unidirectional propagation, lossless information transfer and general immunity to imperfections during fabrication \cite{Ozawa2019}. Hence, in order to fully exploit the advantages that these topological states provide, it is of great experimental interest to devise a method to pump them efficiently while using an input with minimum complexity. To this end, in this work we introduce the idea of adiabatically modifying the geometry and parameters of a discrete system, so that input and output profiles are supersymmetric partners of each other. The framework of Supersymmetry (SUSY), which was first applied to optics in Ref.~\cite{Chumakov1994} and to discrete waveguide lattices in Ref.~\cite{Longhi2010PRB}, has been exploited in recent years to produce applications in modal control \cite{Miri2013,Miri2013A,heinrich2014,principe2015,Macho2018,Queralto2018,Walasik2018,Contreras2019} and Mode-Division Multiplexing \cite{Queralto2017,Walasik2019,Viedma2021}, control of the scattering properties of a system \cite{Longhi2010,heinrich2014OL,Miri2014,Longhi2015,Garcia2020} and its topology \cite{Queralto2020}, among others. 

In lattice systems with nontrivial topology, one can employ discrete SUSY transformations to obtain a superpartner lattice in which the topological state is confined in an isolated site. Performing an excitation on this site, and then a subsequent adiabatic deformation into the original lattice, leads to a faithful pumping of the topological state with minimal excitation of any others. This method can in principle be used to prepare any state of a one-dimensional system, not just topological states, from a single-site excitation, a feat that would otherwise imply simultaneous excitation of multiple sites with the right amplitudes and relative phases.

The method is completely general and does not depend on a particular kind of structure. However, to demonstrate its performance we choose to focus on the geometry of the Su-Schrieffer-Heeger (SSH) model \cite{Su1979}. The SSH chain can host topological edge states, and modified chains with a central defect can also host a topological interface state confined around it. The topological states of the SSH chain have already found applications mainly in contexts like quantum state transfer \cite{Almeida2016,Bello2016,Longhi2019AQT,Longhi2019} and lasing \cite{St-Jean2017,Zhao2018,Parto2018}. We implement the SSH model in an array of single-mode optical waveguides, where the adiabatic deformation can be performed along the propagation direction of light.

The work is organized as follows: In Section \ref{S-Theory}, we outline the main theoretical considerations of the paper, and we introduce the physical system under study in Section \ref{S-System}. Then, we describe the results and applications in Section \ref{S-Results}. Finally, we lay our conclusions in Section \ref{S-Conclusions}.

\section{Theory} \label{S-Theory}

\begin{figure*}[t]
    \centering
    \begin{subfigure}[t]{1.00\textwidth}
    \includegraphics[width=\textwidth]{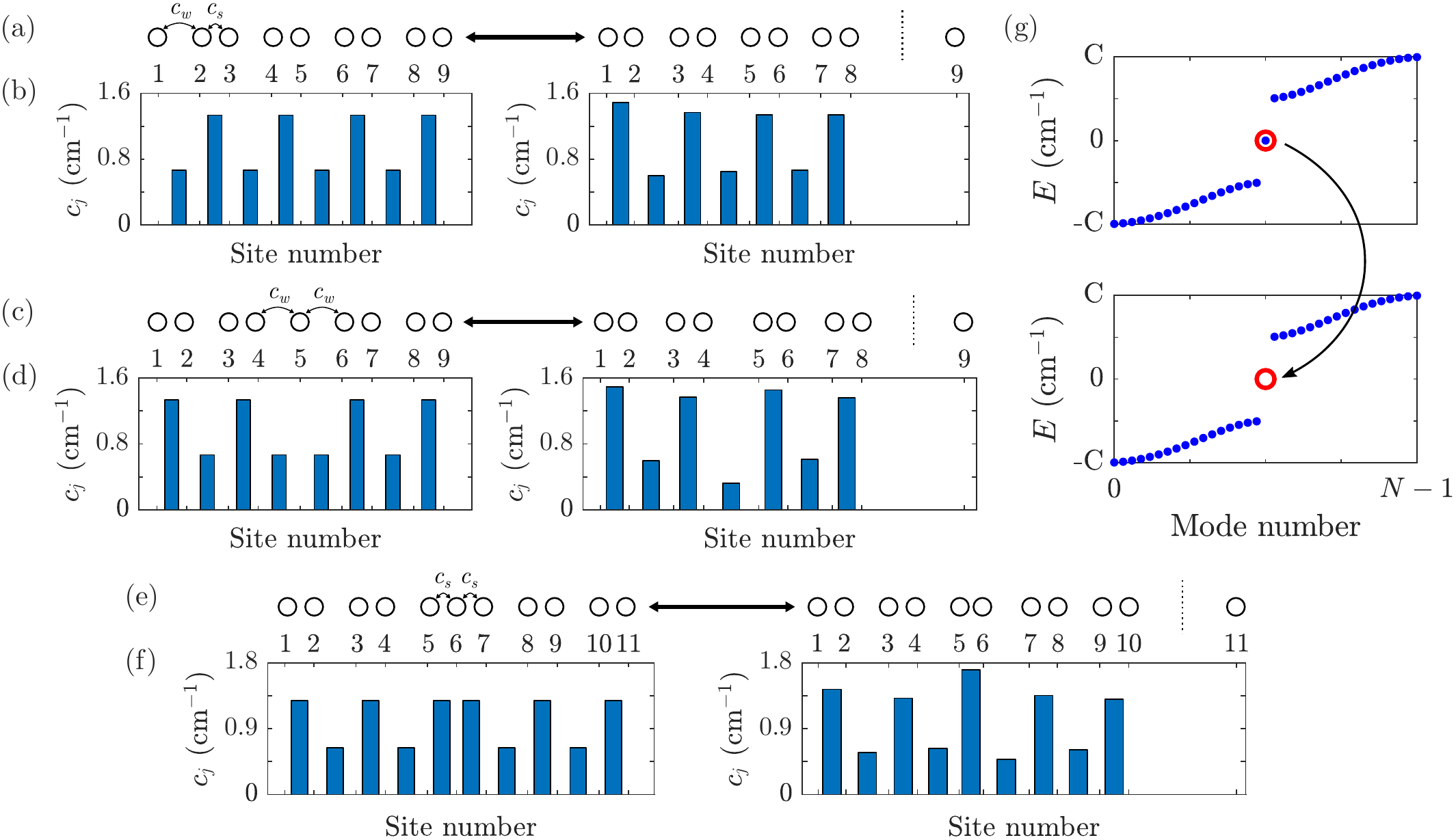}
    \phantomcaption
    \label{fig:edge-sites}
    \end{subfigure}
    \begin{subfigure}[t]{0\textwidth}
    \includegraphics[width=\textwidth]{example-image-b}
    \phantomcaption
    \label{fig:edge-couplings}   
    \end{subfigure}
    \begin{subfigure}[t]{0\textwidth}
    \includegraphics[width=\textwidth]{example-image-b}
    \phantomcaption
    \label{fig:int1-sites}   
    \end{subfigure}
    \begin{subfigure}[t]{0\textwidth}
    \includegraphics[width=\textwidth]{example-image-b}
    \phantomcaption
    \label{fig:int1-couplings}   
    \end{subfigure}
    \begin{subfigure}[t]{0\textwidth}
    \includegraphics[width=\textwidth]{example-image-b}
    \phantomcaption
    \label{fig:int4-sites}   
    \end{subfigure}
    \begin{subfigure}[t]{0\textwidth}
    \includegraphics[width=\textwidth]{example-image-b}
    \phantomcaption
    \label{fig:int4-couplings}   
    \end{subfigure}
    \begin{subfigure}[t]{0\textwidth}
    \includegraphics[width=\textwidth]{example-image-b}
    \phantomcaption
    \label{fig:SSH-spectra}   
    \end{subfigure}
    \caption{(a), (c), (e) Sketches of the configurations of SSH chains (left) and their superpartners (right). The strength of the coupling is represented by the closeness of the sites. The SSH chains can host (a) an edge state localized in the left edge or (c), (e) an interface state localized in the dimer defect. The supersymmetric partner of these modes is localized in the isolated site on the right edge in each case. (b), (d), (f) Structure of the coupling strengths for the chains in (a), (c) and (e), respectively. (g) Spectrum of energies of the SSH chain hosting a single edge state (top) and of the superpartner structure where the eigenvalue is missing (bottom).}
    \label{fig:SSH-sites}
\end{figure*}

Let us consider the Su-Schrieffer-Heeger (SSH) model \cite{Su1979}, which is characterized by two 1D sublattices with staggered weak and strong couplings, $c_w$ and $c_s$ respectively. We display this geometry in the left panel of Fig.~\ref{fig:edge-sites}, where the larger (smaller) the distance between sites, the weaker (stronger) the coupling strength between them. The SSH chain can host topological edge states in one or both of its ends if the outermost coupling is the weak one. The edge states in the SSH model have energies close to zero for finite chains, while for semi-infinite chains their energy is exactly zero. The chiral symmetry of the SSH model ensures that the edge states only have nonzero projections in one of the sublattices \cite{asboth2016}.

Although an SSH chain cannot host a topological interface state, a way to make it appear naturally is to include a dimer defect in a specific position of the chain \cite{Poli2015,Weimann2017,Estarellas2017,Longhi2019}. This kind of defect can be implemented by repeating one of the couplings at a certain position of the staggered configuration, see the left panels of Figs.~\ref{fig:int1-sites} and \ref{fig:int4-sites}, and it is equivalent to connecting two chains with different topological order. In this configuration, the interface state is closely confined around the defect independently of the number of sites of the chain and the values of the couplings, as long as $c_w \neq c_s$. If we include a defect in a topologically nontrivial finite chain, however, the interface state will in general hybridize with the edge states. Additionally, if the defect is generated by the strong coupling $c_s$, additional localized states arise around the defect \cite{Estarellas2017} with energies above and below the two energy bands of the system.

Discrete Supersymmetry (DSUSY) transformations \cite{Miri2013,Longhi2010PRB} can be built between two discrete superpartner Hamiltonians $\mathcal{H}^{(1)}$ and $\mathcal{H}^{(2)}$. Applying these transformations to a discrete system, in general, leads to another system that has a modified geometry -- and thus a different coupling structure -- but shares the same spectrum of energies as the former. However, if the symmetry is unbroken, the superpartner $\mathcal{H}^{(2)}$ displays an isolated site within which the mode with eigenvalue $\mu_m$ targeted by the transformation is confined. The superpartner structures of the SSH model and of the two structures exhibiting interface states are shown in the right panels of Figs.~\ref{fig:edge-sites}, \ref{fig:int1-sites} and \ref{fig:int4-sites}. To obtain the superpartner structures, we employ the QR factorization method \cite{Hogben_2013}. This method allows to remove any desired eigenvalue $\mu_m$, while preserving the rest of the spectrum. The effect of this transformation on the spectrum is displayed in Fig.~\ref{fig:SSH-spectra}, where we have removed the eigenvalue of the edge state of the SSH chain. The QR factorization can be implemented using:
\begin{align}
    \mathcal{H}_m^{(1)}-\mu_m^{(1)} I &= QR  & \text{and}& & \mathcal{H}_m^{(2)}-\mu_m^{(1)} I &= RQ,
    \label{QR-fact}
\end{align}
where $I$ is the identity matrix, $Q$ is an orthogonal matrix and $R$ an upper diagonal matrix. One should note that, in general, this factorization is not unique \cite{Hogben_2013}, and this can lead to more than one superpartner structure sharing the same eigenvalues \cite{Queralto2020}.

QR factorization can be implemented via Gram-Schmidt decomposition, via the Givens rotation method or using the Householder transformation \cite{Hogben_2013}. The last two methods are based on applying unitary transformations to an arbitrary matrix, turning it into a triangular one. Throughout this work we have employed the Householder transformation \cite{Householder1958,Hogben_2013}, which is based on applying successive reflections to each column vector $\bm{x}_i$ of a matrix $A$, via a vector $\bm{u}_i = \bm{x}_i-||\bm{x}_i||\bm{e}_i$ where $\bm{e}_i$ is the unitary vector in the final direction. By defining $Q_i = I - \bm{u}_i\bm{u}_i^T/||\bm{u}_i||^2$, where $^T$ represents the transpose, one can then iteratively build the matrices $Q = \prod_{i=1}^N Q_i^T$ and $R = \left(\prod_{i=1}^N Q_{N-i+1}\right) A$ that lead to $A = QR$. For square matrices, each $Q_i$ acts on the subspace $(N-i+1) \times (N-i+1)$ of $A$, so that $Q_i Q_{i-1}\cdots Q_1 A$ has zeros below the diagonal up to column $i$.

\section{Physical system} \label{S-System}

\begin{figure*}[t]
\centering
\begin{subfigure}{1.00\textwidth}
  \centering
  \includegraphics[width=\linewidth]{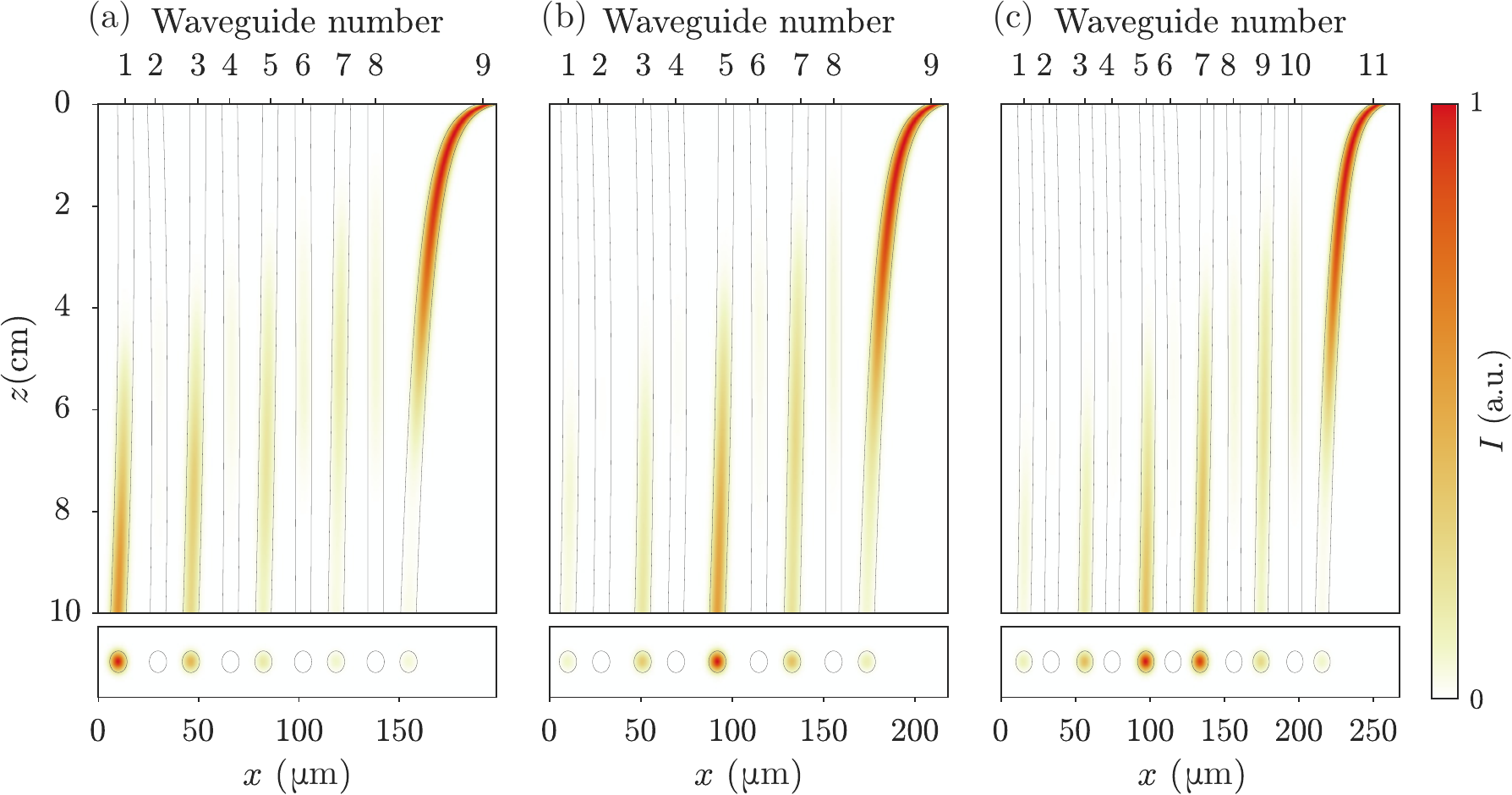}
  \phantomcaption
  \label{fig:prop-edge} 
\end{subfigure}%
\begin{subfigure}{0.0\textwidth}
    \includegraphics[width=0\textwidth]{example-image-b}
    \phantomcaption
    \label{fig:prop-int1}   
\end{subfigure}%
\begin{subfigure}{0\textwidth}
    \includegraphics[width=0\textwidth]{example-image-b}
    \phantomcaption
    \label{fig:prop-int4}   
\end{subfigure}%
\caption{Numerical simulation of the light intensity propagation (top) and final intensity pattern (bottom) in the pumping of (a) the left edge state, (b) the interface state of the weak defect and (c) of the strong defect. Light is injected into the rightmost waveguide, waveguide 9 in (a) and (b), and waveguide 11 in (c) of the SUSY structures.}
\label{fig:prop-SSH}
\end{figure*}

Arrays of optical waveguides in close proximity are coupled due to the overlap of their evanescent fields. If this coupling is not strong enough to affect the mode profiles in each of the waveguides, one can describe arrays of single-mode waveguides using the coupled-mode formalism \cite{Jones1965}, in which light propagation is described by:
\begin{equation}
    -i\frac{d}{dz}\bm{a} = \mathcal{H} \bm{a},
    \label{coupled-mode}
\end{equation}
where $\bm{a} = \left(a_1,\ldots,a_N\right)^T$ are the complex modal amplitudes in each waveguide and $\mathcal{H}$ is a tridiagonal Hamiltonian featuring the propagation constants $\beta_j$ in each waveguide (diagonal) and the couplings $c_{j,j\pm1}$ between them (off-diagonal).

Since the coupling between waveguides is proportional to the overlap of their modes, its strength can be controlled by simply controlling the distance between them, $d$. For each wavelength, the couplings can then be fit into an exponential function \cite{Szameit2007}, $c(d) = c_0  \exp(-\kappa d)$, and after obtaining $c_0$ and $\kappa$ one can perform a mapping between couplings and distances. For DSUSY transformations targeting a zero-energy state, the propagation constants are not altered. Therefore, for this work, it will be sufficient to ensure that all waveguides share the same propagation constant.

In practice, if one builds the superpartner lattice of a particular system -- which retains the nearest-neighbor couplings configuration -- one obtains supermodes that are phase-matched to the originals and have the same eigenvalues, but where the mode corresponding to the removed eigenvalue is localized in an isolated waveguide. By exciting this localized mode, and adiabatically deforming the SUSY lattice into the SSH one, we can obtain the mode that was removed with high fidelity, starting from a single-waveguide excitation.

The adiabatic connection between the SUSY and SSH structures can be performed in multiple ways, as multiple paths can be followed in the hyperspace of parameters. We have chosen to follow paths in which every parameter evolves with the same type of transformation function, $g(z)$:
\begin{equation}
    \bm{c}(z) = \bm{c}_{SUSY} + g(z)\left(\bm{c}_{SSH}-\bm{c}_{SUSY}\right),
    \label{transf-function-c}
\end{equation}
where $\bm{c}_i = \left(c_{i,1},\ldots,c_{i,N}\right)$ with $i = SSH, SUSY$ are the couplings in the input and output structures, respectively. For each case in question, one could engineer the transformation function that optimizes the adiabatic pumping. However, a single function will not optimize the method for all parameters. Since the method itself does not depend on the particular path along coupling space, we choose to use a simple linear function in $z$ for all cases in this work. Despite that, one should keep in mind that results can be further improved by performing this optimization process.

A linear transformation in the couplings implies the following dependence for the distance between waveguides $j$ and $j+1$:
\begin{equation}
    d_j(\tilde{z}) = \frac{1}{\kappa}\log\left(\frac{c_0}{(1-\tilde{z})\,c_{SUSY,j} + \tilde{z}\, c_{SSH,j}}\right),
    \label{eq-distances}
\end{equation}
where $\tilde{z}=z/L$ is the coordinate along the propagation direction scaled to the length of the device, $L$. To compute the distances we have considered state-of-the-art experimental parameters for a wavelength of $\lambda = 633$ nm for laser-written waveguides \cite{Szameit2010,Queralto2020}.

Since this method relies on the adiabaticity of the transformation, it will benefit fidelity-wise from a longer setup. The general adiabaticity condition when following the state $\psi_m$ reads: $\left|\left<\psi_l|\partial_z\psi_m\right>\right| \ll \left|\beta_m-\beta_l\right|$ for all $l$, that is, the coupling between the instantaneous eigenstates $\psi_{l}$ and $\psi_{m}$ has to be small compared to the difference between their eigenvalues, $\beta_{m}-\beta_{l}$. In other words, to keep nonadiabatic effects to a minimum, the length of the device has to be large enough for the coupling to be much smaller than the minimum eigenvalue gap.

\section{Results} \label{S-Results}

\begin{figure}[t]
    \centering
    \begin{subfigure}[t]{0.70\columnwidth}
    \includegraphics[width=\linewidth]{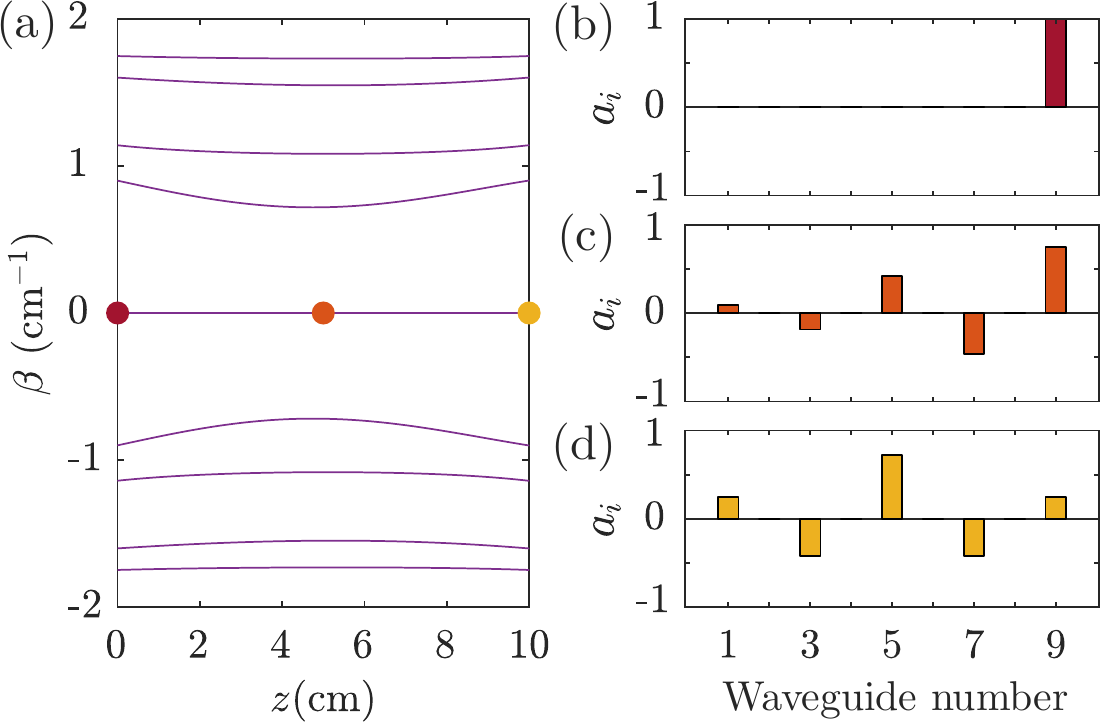}
    \phantomcaption
    \label{fig:spectra-z}
    \end{subfigure}
    \begin{subfigure}[t]{0\textwidth}
    \includegraphics[width=0\textwidth]{example-image-b}
    \phantomcaption
    \label{fig:profile_0}   
    \end{subfigure}
    \begin{subfigure}[t]{0\textwidth}
    \includegraphics[width=0\textwidth]{example-image-b}
    \phantomcaption
    \label{fig:profile_5}   
    \end{subfigure}
    \begin{subfigure}[t]{0\textwidth}
    \includegraphics[width=0\textwidth]{example-image-b}
    \phantomcaption
    \label{fig:profile_10}   
    \end{subfigure}
    \caption{(a) Spectrum of propagation constants along the propagation direction of the device displayed in Fig.~\ref{fig:prop-int1}. (b), (c), (d) Amplitude profile of the target mode at (a) $z = 0$, (b) $z = L/2$, (c) $z = L$ where $L$ is the total length of the device. The positions and propagation constants corresponding to plots (b), (c) and (d) are marked in (a) using dots with the same colors.}
    \label{fig:spectra_profile}
\end{figure}

We have analysed three different devices, in which we connect the configurations displayed in the left and right panels of Fig.~\ref{fig:edge-sites}, \ref{fig:int1-sites} and \ref{fig:int4-sites}, respectively, by adiabatically deforming one waveguide lattice into the other along the propagation direction. In each case, the input configuration is the SUSY partner (right panels) and the output configuration is the corresponding SSH chain (left panels). In all three cases, we excite the isolated waveguide of the input configuration, and we obtain a topological state in the output chain. Namely, we pump the edge state of the SSH chain with the first device (Fig.~\ref{fig:edge-sites}) and the interface states spawned by dimer defects with the latter devices (Figs.~\ref{fig:int1-sites} and \ref{fig:int4-sites}). Throughout the work, we refer to these devices as adiabatic SUSY (aSUSY) devices. As an example, we sketch in Fig.~\ref{fig:spectra-z} the spectra of the device displayed in Fig.~\ref{fig:prop-int1} along $z$. During the propagation, the single-waveguide excitation adiabatically transforms into the target mode, as shown in Figs.~\ref{fig:profile_0}--\ref{fig:profile_10}.

To characterize the efficiency of each of the designed devices, we define the fidelity:
\begin{equation}
    F = \left|\left<\phi_{k}|\psi(z_f)\right>\right|^2,
    \label{fid}
\end{equation}
where $\phi_{k}$ is the SSH state we aim to produce and $\psi(z_f)$ is the output state obtained from the numerical simulations.

\subsection{Edge state}

\begin{figure}[t]
    \centering
    \includegraphics[width=0.5\linewidth]{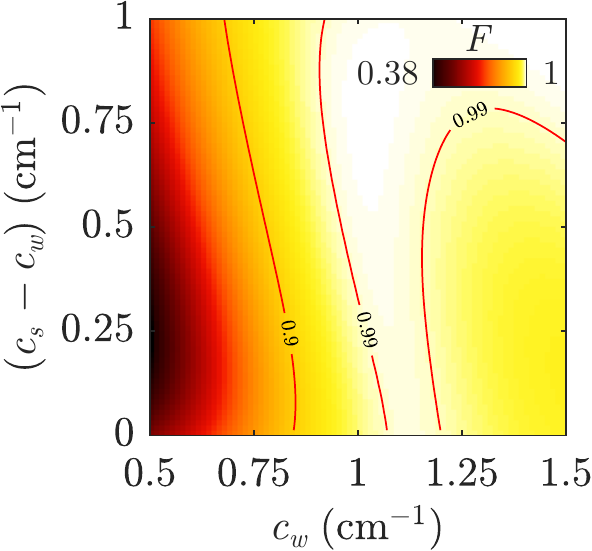}
    \caption{Fidelity of the pumping of the left edge state, for an aSUSY device of length $L=\SI{10}{\cm}$ and different values of the SSH couplings. We maintain $c_s > c_w$ for all considered values, and indicate the regions where the fidelity exceeds $F=0.9$ and $F=0.99$ with solid red lines.}
    \label{fig:f-edge}
\end{figure}

\begin{figure}[t]
    \centering
    \begin{subfigure}[t]{0.70\columnwidth}
    \includegraphics[width=\linewidth]{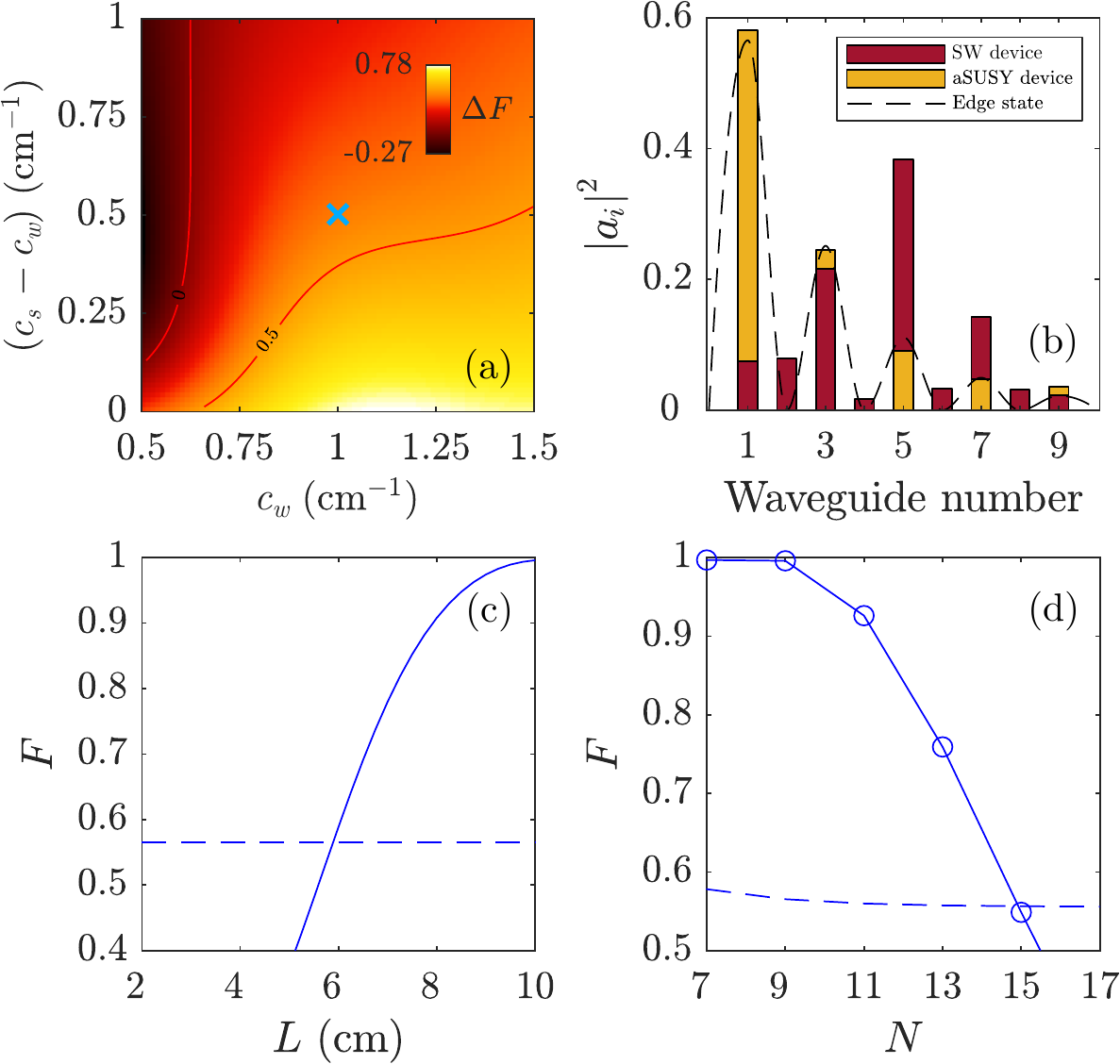}
    \phantomcaption
    \label{fig:df-edge}
    \end{subfigure}
    \begin{subfigure}[t]{0\textwidth}
    \includegraphics[width=0\textwidth]{example-image-b}
    \phantomcaption
    \label{fig:comparison-edge}   
    \end{subfigure}
    \begin{subfigure}[t]{0\textwidth}
    \includegraphics[width=0\textwidth]{example-image-b}
    \phantomcaption
    \label{fig:f_d-edge}   
    \end{subfigure}
    \begin{subfigure}[t]{0\textwidth}
    \includegraphics[width=0\textwidth]{example-image-b}
    \phantomcaption
    \label{fig:f_N-edge}   
    \end{subfigure}
    \caption{(a) Difference in fidelity $\Delta F = F - F_{s}$ between the aSUSY device of length $L=10$ cm and the SW device. (b) Comparison between the squared amplitudes of the state generated using the aSUSY device in yellow and the SW device in red for $L=10$ cm. The profile of the actual edge state is depicted with a dashed line. (c) Fidelity of the aSUSY (solid) and of the SW (dashed) devices of $N=9$ for increasing device length. (d) Fidelity of the aSUSY (circles) and of the SW (dashed line) devices of length $L=\SI{10}{\cm}$ for increasing number of waveguides. The value of the couplings for (b), (c) and (d) corresponds to the light blue cross in (a), $c_w = \SI{1}{\cm^{-1}}$ and $c_s = \SI{1.5}{\cm^{-1}}$.}
    \label{fig:imp-edge}
\end{figure}

\begin{figure*}[t]
\centering
\begin{subfigure}{1\textwidth}
    \centering
    \includegraphics[width=1.0\linewidth]{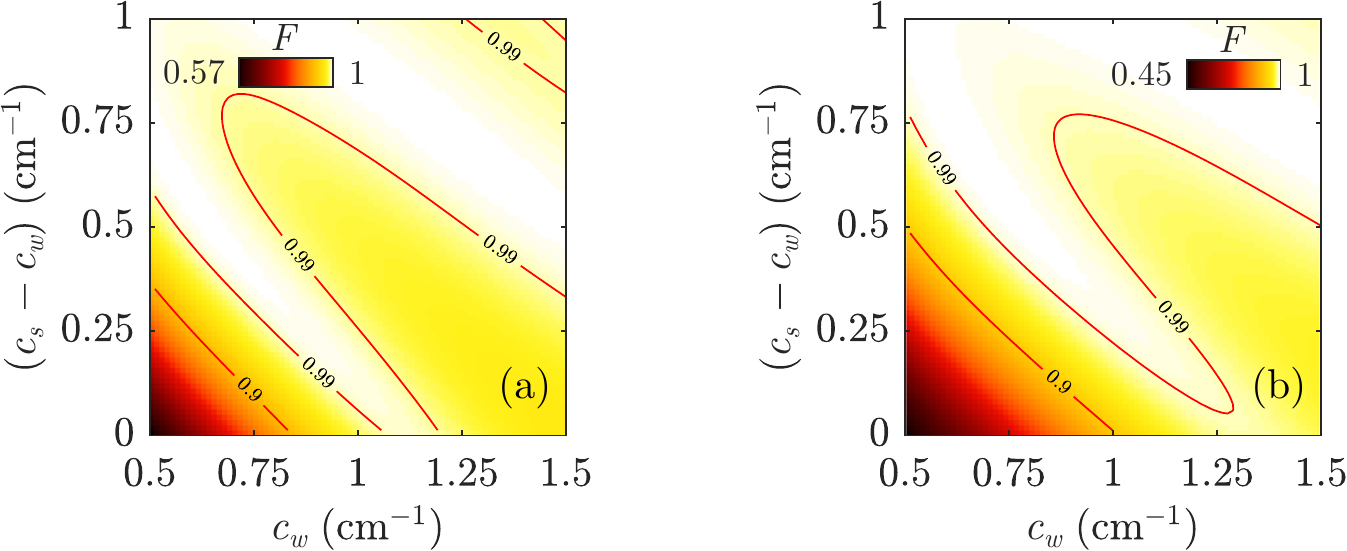}
    \phantomcaption
    \label{fig:f_int1}
\end{subfigure}%
\begin{subfigure}{0\textwidth}
    \centering
    \includegraphics[width=0.0\linewidth]{example-image-b}
    \phantomcaption
    \label{fig:f_int4}
\end{subfigure}
\caption{Fidelity of the pumping of the interface state in the defect formed by repeating (a) the weak coupling and (b) the strong coupling for an aSUSY device of length $L=\SI{10}{\cm}$. We maintain $c_s > c_w$ for all considered values, and indicate the regions where the fidelity exceeds $F=0.9$ and $F=0.99$ with solid red lines.}
\label{fig:f-int-def}
\end{figure*}

We first focus on the pumping of the left edge state of the SSH lattice corresponding to the left panel of Fig.~\ref{fig:edge-sites}, starting from the excitation of the rightmost waveguide of the SUSY partner depicted in the right panel of Fig.~\ref{fig:edge-sites}. The distances between waveguides are determined using Eq.~(\ref{eq-distances}). We display in Fig.~\ref{fig:prop-edge} the propagation of the excitation along the device, and the intensity of the mode at the output facet. The obtained mode clearly resembles the edge state, being localized around the leftmost waveguide and having non-zero amplitude only in one of the sublattices of the chain.

Let us now focus on how efficiently we obtain the edge state in the aSUSY device. In Fig.~\ref{fig:f-edge} we scan the two-dimensional (2D) space of couplings, maintaining $c_s>c_w$ so that we do not change the topological order of the chain, and we plot the fidelity for a device of length $L = \SI{10}{\cm}$. We observe that as long as $c_w \gtrsim \SI{0.8}{\cm^{-1}}$, the fidelities we obtain are remarkably good, with values surpassing $F=0.9$ for most parameters and with a wide region where $F>0.99$. In contrast, if $c_w$ is smaller than that value, the injected mode requires longer devices to be efficiently transformed into the edge state at the other end of the chain. Energy-wise, this decrease in efficiency is due to a decrease of the energy gap, which implies that slower deformations are required to maintain the adiabaticity of the transformation. The appearance of nonadiabatic couplings implies the loss of power to other modes.

The fidelity of the aSUSY device should be compared with the standard injection of light at the outermost waveguide -- position at which the edge state has the largest amplitude -- of an SSH lattice with parallel waveguides of the same length as the aSUSY device. We refer to the device with this input as the \textit{straight waveguides} (SW) device. For this case, the input will in general have a large projection over the edge state, but will also excite other nonlocalized modes. 

In Fig.~\ref{fig:df-edge} we plot $\Delta F = F - F_{s}$, where $F$ is the fidelity of the aSUSY device and $F_{s}$ is that of the SW device. The improvement of the former with respect to the latter will depend on how closely confined the edge state is within the outermost waveguide of the SSH lattice, which in turn depends on the ratio between couplings. When the ratio $c_s/c_w$ is close to one, as long as $c_w \gtrsim \SI{0.8}{\cm^{-1}}$, our technique allows us to reach the edge state almost perfectly, whereas the SW output has only a small projection over it. In Fig.~\ref{fig:comparison-edge} we show the comparison between the intensity profiles of the real edge state and both the state generated by the adiabatic deformation of the structure and the state achieved by letting the input propagate through the SW device. From this comparison, we see that the adiabatic modification of the lattice allows to correctly capture the decaying amplitude of the mode. If we extend the device while maintaining the geometry of the SSH chain, the intensity pattern of the mode that we obtain with the aSUSY device will remain mostly unaltered after a certain propagation distance, while the input in the SW device disperses due to the excitation of nonlocalized modes. This effect is more apparent the less confined the edge mode is.

Two questions now arise regarding the size of the devices, how small in length and how big in number of waveguides can we make them. The answers will depend on all the parameters of the system, the mode we are trying to pump and the fidelity that is needed. We show in Fig.~\ref{fig:f_d-edge} the increase in fidelity as we increase the length of the aSUSY device (solid line). Comparing it with the SW device (dashed line), we see that the adiabatic modification starts being more efficient after the device length reaches a few centimeters, around $L=6$ cm. On the other hand, we see in Fig.~\ref{fig:f_N-edge} that the device has higher fidelities for lower number of waveguides, due to the energy gap getting smaller as $N$ increases. Comparing with the SW device, for $L = 10$ cm our method to obtain the edge state stops being efficient at around $N = 15$ for this set of couplings. Logically, the device would need larger couplings or longer distances for light intensity to traverse more waveguides.

\begin{figure}[t]
    \centering
    \begin{subfigure}[t]{0.70\columnwidth}
    \includegraphics[width=\linewidth]{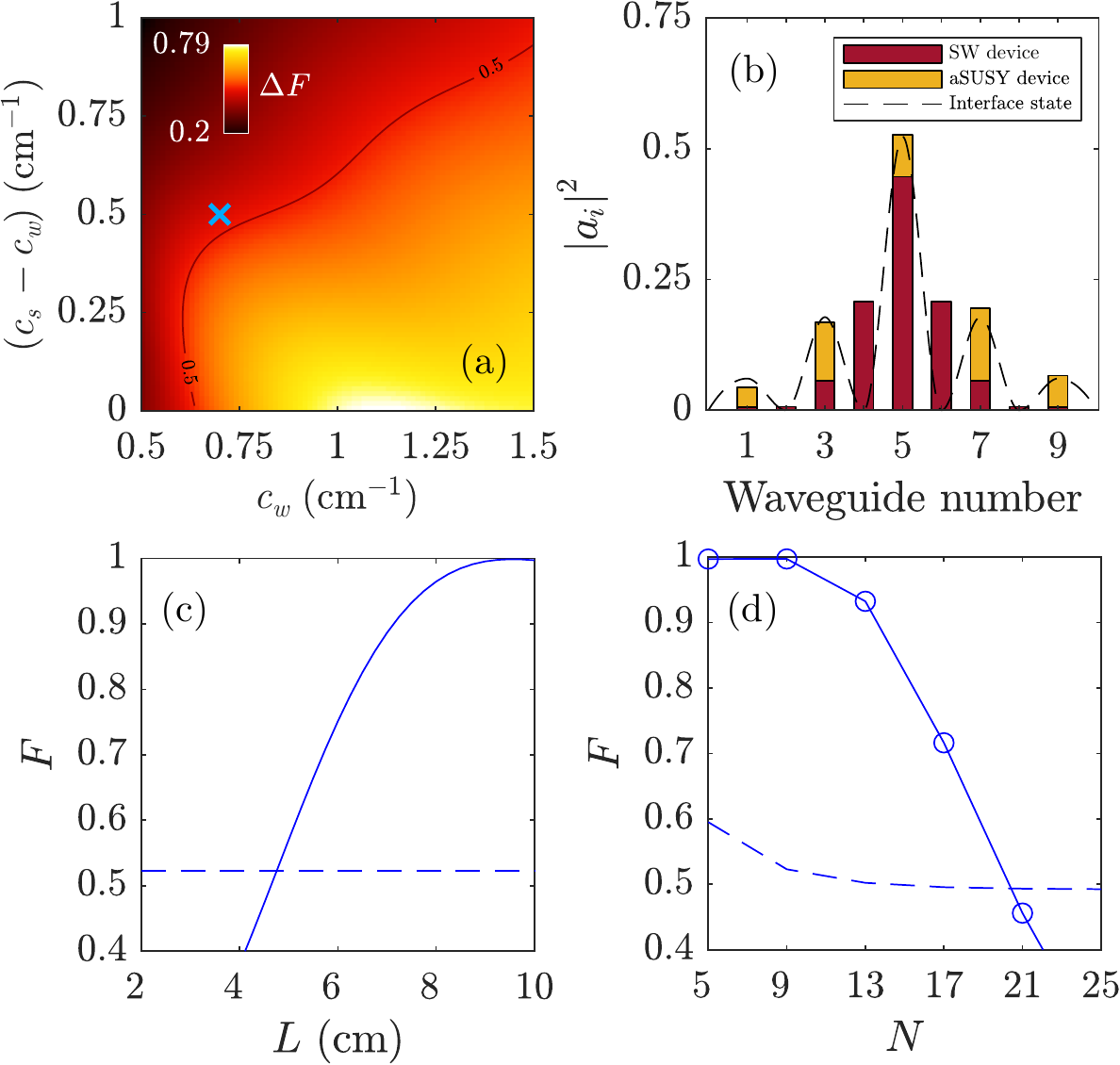}
    \phantomcaption
    \label{fig:df-int1}
    \end{subfigure}
    \begin{subfigure}[t]{0\textwidth}
    \includegraphics[width=0\textwidth]{example-image-b}
    \phantomcaption
    \label{fig:comparison-int1}   
    \end{subfigure}
    \begin{subfigure}[t]{0\textwidth}
    \includegraphics[width=0\textwidth]{example-image-b}
    \phantomcaption
    \label{fig:f_d-int1}   
    \end{subfigure}
    \begin{subfigure}[t]{0\textwidth}
    \includegraphics[width=0\textwidth]{example-image-b}
    \phantomcaption
    \label{fig:f_N-int1}   
    \end{subfigure}
    \caption{(a) Difference in fidelity $\Delta F = F - F_{s}$ between the aSUSY device of length $L=10$ cm and the SW device when pumping the interface state corresponding to the weak coupling defect. (b) Comparison between the squared amplitudes of the state generated by the aSUSY device in yellow and the SW device in red for $L=10$ cm. The profile of the actual interface state is depicted with a dashed line. (c) Fidelity of the aSUSY (solid) and of the SW (dashed) devices of $N=9$ for increasing device length. (d) Fidelity of the aSUSY (circles) and of the SW (dashed line) devices of length $L=10$ cm for increasing number of waveguides. The value of the couplings for (b), (c) and (d) corresponds to the light blue cross in (a), $c_w = \SI{0.7}{\cm^{-1}}$ and $c_s = \SI{1.2}{\cm^{-1}}$.}
    \label{fig:imp-int1}
\end{figure}

\begin{figure}[t]
    \centering
    \begin{subfigure}[t]{0.70\columnwidth}
    \includegraphics[width=\linewidth]{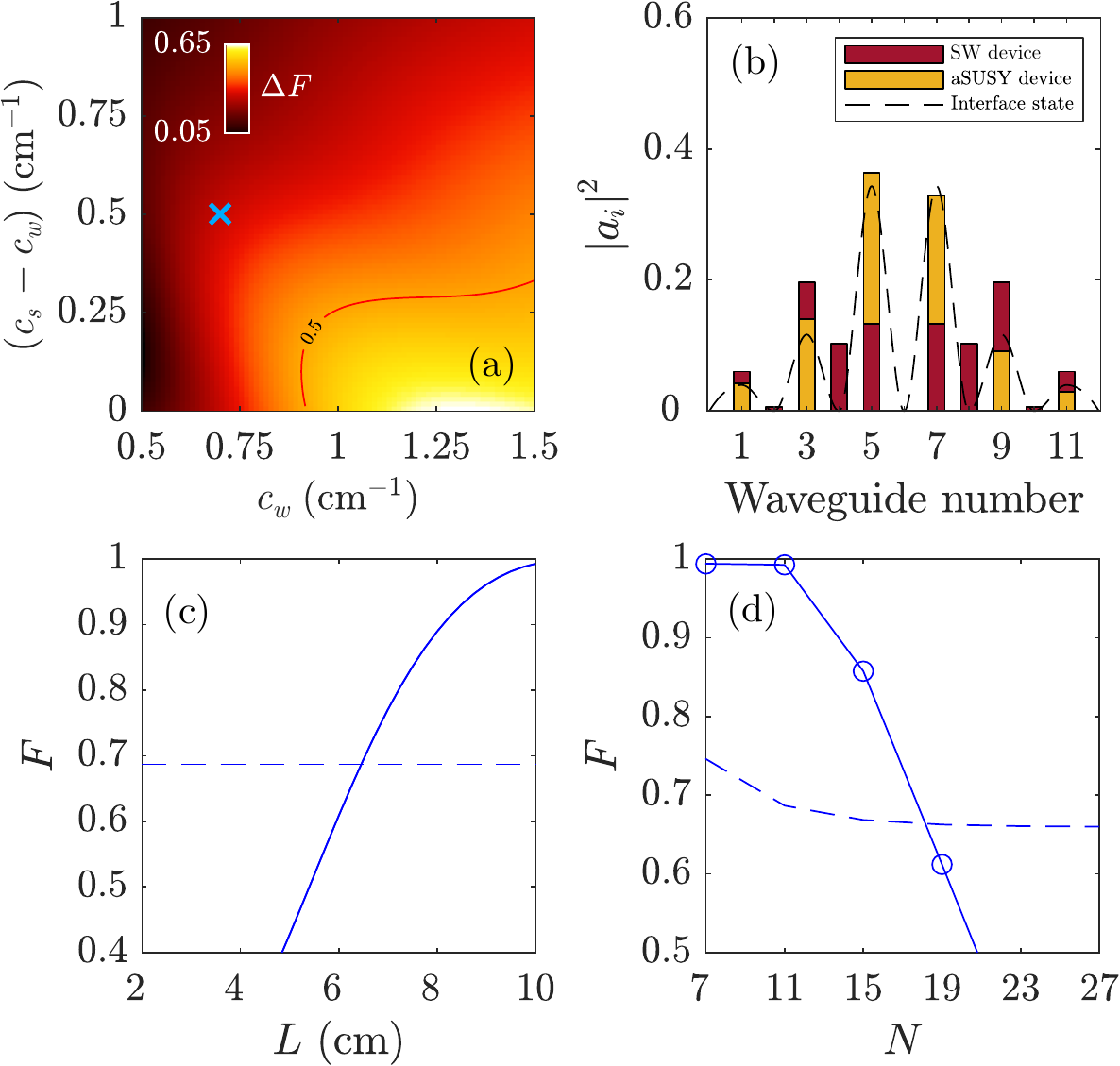}
    \phantomcaption
    \label{fig:df-int4}
    \end{subfigure}
    \begin{subfigure}[t]{0\textwidth}
    \includegraphics[width=0\textwidth]{example-image-b}
    \phantomcaption
    \label{fig:comparison-int4}   
    \end{subfigure}
    \begin{subfigure}[t]{0\textwidth}
    \includegraphics[width=0\textwidth]{example-image-b}
    \phantomcaption
    \label{fig:f_d-int4}
    \end{subfigure}
    \begin{subfigure}[t]{0\textwidth}
    \includegraphics[width=0\textwidth]{example-image-b}
    \phantomcaption
    \label{fig:f_N-int4}   
    \end{subfigure}
    \caption{(a) Difference in fidelity $\Delta F = F - F_{s}$ between the aSUSY device of length $L=10$ cm and the SW device when pumping the interface state generated by the strong coupling defect. (b) Comparison between the squared amplitudes of the state generated by the aSUSY device in yellow and the SW device in red for $L=10$ cm. The profile of the actual interface state is depicted with a dashed line. (c) Fidelity of the aSUSY (solid) and of the SW (dashed) devices of $N=11$ for increasing device length. (d) Fidelity of the aSUSY (circles) and of the SW (dashed line) devices of length $L=10$ cm for increasing number of waveguides. The value of the couplings for (b), (c) and (d) corresponds to the light blue cross in (a), $c_w = \SI{0.7}{\cm^{-1}}$ and $c_s = \SI{1.2}{\cm^{-1}}$.}
    \label{fig:imp-int4}
\end{figure}

\subsection{Interface state}

We now focus on the interface states generated by dimer defects in otherwise topologically trivial SSH chains (left panels of Figs.~\ref{fig:int1-sites} and \ref{fig:int4-sites}). The pumping again starts from a single-waveguide excitation of the isolated waveguide of their SUSY partners. The propagation of these excitations through the devices and the respective output states are shown in Figs.~\ref{fig:prop-int1} and \ref{fig:prop-int4}. In the former case, where the defect is spawned by the weak coupling, we see that the interface state is mostly confined within the central waveguide (waveguide 5 in Fig.~\ref{fig:prop-int1}). Conversely, the maximum of light intensity splits into two waveguides (waveguides 5 and 7 in Fig.~\ref{fig:prop-int4}) when the defect is formed by repeating the strong coupling. From these figures, it is also clear that the interface states only have non-zero projections over one of the sublattices, just like the edge states of the defectless SSH chain.

The efficiencies when pumping these two states are remarkably good, as can be observed from Figs.~\ref{fig:f_int1} and \ref{fig:f_int4}, respectively. In both figures, the fidelity surpasses $F=0.99$ in most of the parameter space that we consider. Both results are fairly similar, but some differences can be spotted in the map of fidelities. The overall values are slightly smaller in Fig.~\ref{fig:f_int4}, and the patterns are displaced. These differences are mainly due to the different total number of waveguides considered for each case. Each device needs to have a different number of waveguides if we want to maintain the topological order of the rest of the SSH chain whilst adding the two different defects at the center.

As before, we compare these results with the respective SW devices. For the pumping of the interface state formed by the weak coupling defect, which is displayed in Fig.~\ref{fig:prop-int1}, the input of the SW device will consist in the injection of light in waveguide 5 of the chain of straight waveguides. We display the difference in fidelity between the devices in Fig.~\ref{fig:df-int1}, where we see that the biggest improvement achieved by the aSUSY device is obtained in regions where the dimerization $|c_s-c_w|$ is small, especially around $|c_s-c_w|\le 0.5$ cm$^{-1}$. In a similar way as for the case of the edge state, the adiabatic procedure remains efficient even if the interface state is not strongly confined, while the SW device fails to precisely reproduce the state. The comparison between the intensity patterns of the output state of both devices with the actual interface state is shown in Fig.~\ref{fig:comparison-int1}.

As for the pumping in Fig.~\ref{fig:prop-int4}, corresponding to the defect formed by the strong coupling, there is no simple way to obtain a state that is similar to the interface state, since its maximum of intensity is split into two waveguides, and there is a phase difference between them. The output of the aSUSY device correctly captures these properties, while a SW device with a single-waveguide excitation simply cannot.
To make a fair comparison, we assume that we can perfectly produce a split beam with the right phases beforehand, and that we use it as the input of the SW device. This corresponds to the excitation of waveguides 5 and 7. We make the comparison of fidelities and of intensity profiles with this device in Figs.~\ref{fig:df-int4} and \ref{fig:comparison-int4}, respectively. In those, we readily see that the results are similar to the ones for the previous interface state, which is remarkable considering the increased complexity of the SW device.

We now compare the fidelity of the aSUSY and SW devices for increasing device length and number of waveguides for each case. For the first interface state, as shown in Figs.~\ref{fig:f_d-int1} and \ref{fig:f_N-int1}, the device starts being efficient when its length is close to $L=5$ cm, and maintains higher efficiencies for $L=10$ cm until the number of waveguides reaches $N=21$. The results for the second interface state plotted in Figs.~\ref{fig:f_d-int4} and \ref{fig:f_N-int4} are similar. In this case, the aSUSY device outperforms the SW device shortly after reaching $L=6$ cm, and keeps being superior for any number of waveguides below $N=29$ for the same set of couplings as the previous case. Comparing both of them with the results for the edge state displayed in Fig.~\ref{fig:imp-edge} in the previous section, we see a sizable improvement in this case, as the aSUSY devices pump the desired modes efficiently in shorter lengths and with larger $N$.

\subsection{Quasi-degenerate and bulk states}

\begin{figure}[t]
    \centering
    \begin{subfigure}[t]{0.70\columnwidth}
    \includegraphics[width=\linewidth]{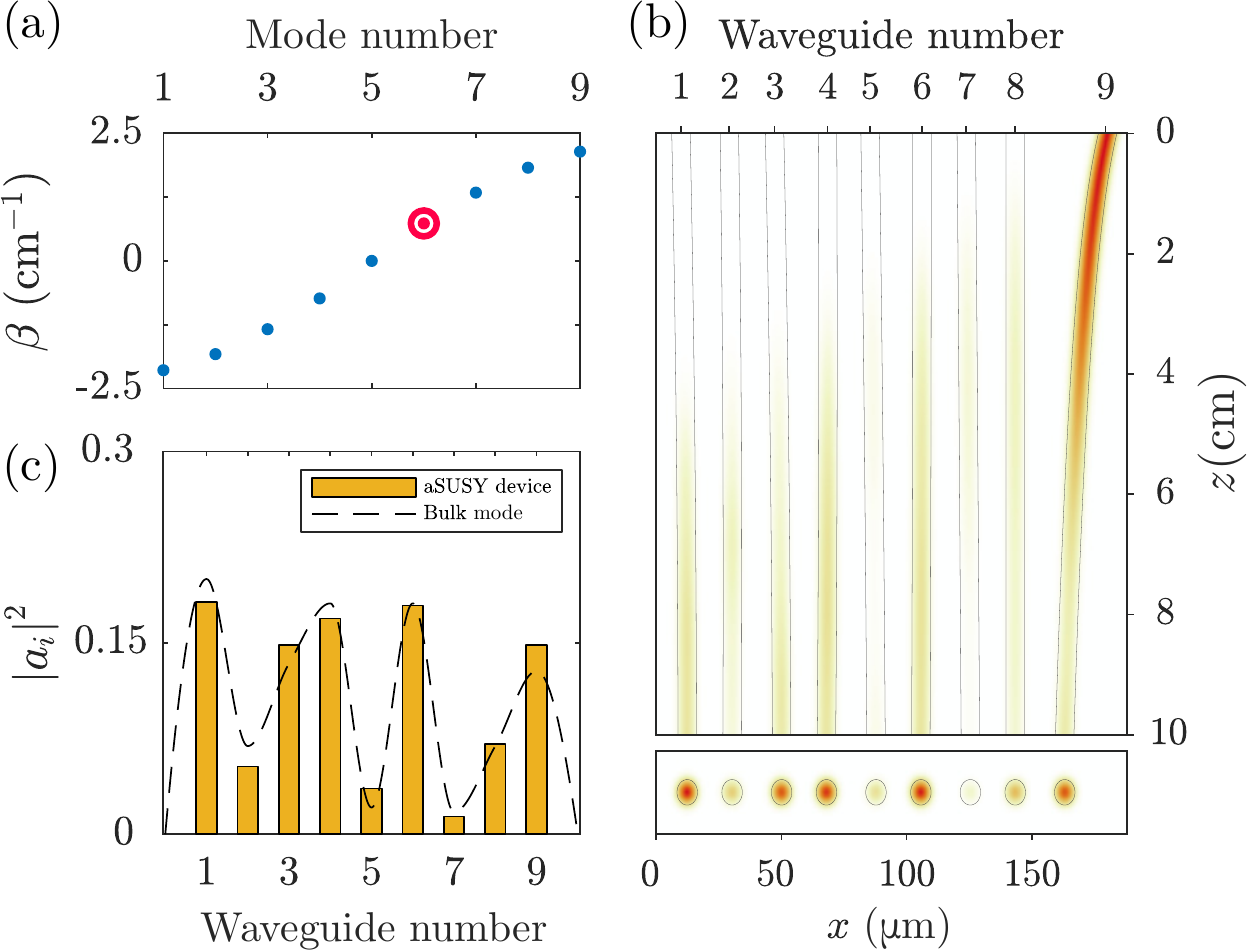}
    \phantomcaption
    \label{fig:spectra-bulk1}
    \end{subfigure}
    \begin{subfigure}[t]{0\textwidth}
    \includegraphics[width=0\textwidth]{example-image-b}
    \phantomcaption
    \label{fig:propagation-bulk1}   
    \end{subfigure}
    \begin{subfigure}[t]{0\textwidth}
    \includegraphics[width=0\textwidth]{example-image-b}
    \phantomcaption
    \label{fig:profile-bulk1}
    \end{subfigure}
    \caption{(a) Spectrum of propagation constants of the SSH model with one edge mode. The bulk mode under consideration is highlighted in red. (b) Intensity propagation for the pumping of the considered bulk state in the SSH model. (c) Output profile (yellow blocks) and target mode (dashed line) for the device in (b). The values for the couplings for (a), (b) and (c) are $c_w = \SI{1}{cm^{-1}}$ and $c_s = \SI{1.5}{cm^{-1}}$.}
    \label{fig:bulk}
\end{figure}

We emphasize that the presented method can be used to pump any mode of the original lattice, not only localized topological modes. In principle, one can target any mode and obtain a superpartner lattice from which to start the pumping from a single-waveguide excitation. As an example, we can consider the bulk modes of the SSH model. We see in Fig~\ref{fig:SSH-spectra} that the gap between bulk modes is smaller than the gap between edge and bulk, which means that generally longer devices are needed to achieve efficient pumping. Nonetheless, for low number of waveguides, bulk modes can be reached within the lengths considered in this work. Let us consider a SSH chain of $N=9$, whose spectra is shown in Fig.~\ref{fig:spectra-bulk1}, and attempt to pump the bulk mode that is highlighted in red. We show in Fig.~\ref{fig:propagation-bulk1} the light intensity propagation through the device. As we can see, unlike the cases discussed previously in this work, the output mode is completely delocalized. In Fig.~\ref{fig:profile-bulk1} we compare the squared amplitudes of the output mode with the target bulk mode, and see that they are remarkably similar. For the parameters used in the simulation, we compute the fidelity to be $F = \num{0.992}$. The downside to pumping bulk modes is that the input structure no longer displays zero detunings. As such, control over the propagation constants of the individual waveguides along the propagation direction is required. This can be achieved in laser-writing setups by controlling the relative speed of the writing laser and the laser \cite{Szameit2007}.

For degenerate or quasi-degenerate states, there is no way to guarantee the absence of nonadiabatic couplings. Therefore, in general, the output mode will be a superposition of the degenerate modes and not a single mode of the lattice. This effect can be readily seen in an SSH chain with two edge modes, corresponding to symmetric and antisymmetric combinations of left and right edges. These are nearly degenerate, so when trying to obtain one of them, a certain fraction of power will inevitably be transferred to the other. This can be somewhat exploited to obtain a certain proportion of both edges as the output state, which will depend on the length of the device and the couplings that are chosen.

\section{Conclusions} \label{S-Conclusions}

In this work, we have shown that by adiabatically connecting a structure with its SUSY partner, we can efficiently transform a single-site excitation into any gapped mode of the structure under investigation. We have focused on the topological modes of the SSH model and we have proposed the implementation of the method using optical waveguide lattices as the physical system. This allows us to perform the adiabatic deformation in space, along the propagation direction of the waveguides. However, the method can be extended to discrete systems with gapped energy levels in any physical platform.

For both the edge state of an SSH lattice and the interface states of SSH lattices featuring a dimer defect, we have obtained fidelities above $F=0.99$ for a wide range of parameters of the system, with optical devices of the order of $L=10$ cm of length. These fidelities are much larger than the ones obtained by direct excitation of the waveguides where the target modes have the largest amplitude, implying a more precise excitation of the relevant mode and a reduction of intensity beatings. The adiabatic modification proposed here is specially useful if the desired output state is not strongly confined, as direct excitation of such states can be extremely hard to accomplish. The energy gap between the target mode and the rest directly determines the fidelity of the method. As such, devices with larger number of waveguides need to be longer to remain efficient. 

The strength of the proposed method lies both in the generality of the modes that can be pumped and the simplicity of the input that is required to pump them. Although some of the modes displayed in this work could be pumped with simpler adiabatic schemes, the DSUSY formalism allows to obtain both localized and nonlocalized modes in general lattices without any added complexity needed for the input beam. Additionally, DSUSY can be used to build other kinds of devices. One could for instance build a device converting an edge mode to an interface mode. This is based on the fact that the QR factorization is not unique, and thus there can exist more than one supersymmetric partner for the same mode of the SSH chain: one supporting an edge state, and one supporting an interface state \cite{Queralto2020}. The device could be built by connecting these two structures adiabatically. DSUSY transformations can also be used in 2D systems \cite{Walasik2018,Smirnova2019,Zhong2019}. This opens new possibilities to generalize the high-efficiency topological pumping method discussed here to higher-dimensional settings, although the superpartner structure of a 2D lattice may host nonlocal couplings, which are more challenging to implement physically.

\begin{backmatter}

\bmsection{Funding} Ministerio de Ciencia, Innovación y Universidades (FIS2017-86530-P, PID2020-118153GB-I00); Generalitat de Catalunya (SGR2017-1646).

%The authors acknowledge financial support from the Spanish Ministry of Science and Innovation MICINN (contracts no. FIS2017-86530-P and PID2020-118153GB-I00) and Generalitat de Catalunya (contract no. SGR2017-1646).

\bmsection{Disclosures} The authors declare no conflicts of interest.

\bmsection{Data availability} Data underlying the results presented in this paper are not publicly available at this time but may be obtained from the authors upon reasonable request.

\end{backmatter}

%%%%%%%%%%%%%%%%%%%%%%% References %%%%%%%%%%%%%%%%%%%%%%%%%

%%%%%%%%%% If using BibTeX:
\bibliography{biblio}

\end{document}